\documentclass[12pt]{article}
\usepackage{epsfig}
\begin{document}
\title{Isospin Effect on the Process of Multifragmentation and
Dissipation at Intermediate Energy Heavy Ion Collisions}
\author{{\small Jian-Ye Liu$^{1,2,3}$, Yan-Fang Yang$^{3}$,
 Wei Zuo$^{1,2,3}$}\\
{\small Shun-Jin Wang$^{1,2,4}$, Qiang Zhao$^{3}$, Wen-Jun Guo$^{3}$,
 Bo Chen$^{3}$}}
\date{}
\maketitle
\begin{center}
$^{1}${\small CCAST (World Lab.), P.O.Box 8730, Beijing 100080}\\
$^{2}${\small Center of Theoretical Nuclear Physics,National
Laboratory of Heavy Ion Accelerator}\\
{\small Lanzhou 730000, P. R. China}\\
$^{3}${\small Institute of Modern Physics, Chinese Academy of
Sciences, P. O. Box 31}\\
{\small Lanzhou 730000, P. R. China}\\
$^{4}${\small Institute of Modern Physics, southwest
Jiaotong University, Chendu 610031, P. R. China}\\
\end{center}
\begin{center}
\begin{minipage}{140mm}
\baselineskip 0.2in
\begin{center}{\bf Abstract}\end{center}
\small
\hskip 0.3in
In the simulation of intermediate energy heavy ion collisions
by using the isospin dependent quantum molecular dynamics, the
isospin effect on the process of multifragmentation and dissipation has
been studied. It is found that the multiplicity of intermediate mass
fragments $N_{imf}$ for the neutron-poor colliding system is always
larger than that for the neutron-rich system, while the quadrupole of
single particle momentum distribution $Q_{zz}$ for the neutron-poor
colliding system is smaller than that of the neutron-rich system for
all projectile-target combinations studied at the beam energies
from about 50MeV/nucleon to 150MeV/nucleon.
Since $Q_{zz}$ depends strongly on isospin dependence
of in-medium nucleon-nucleon cross section and weakly on symmetry
potential at the above beam energies,
it may serve as a good probe to extract the information on the
in-medium nucleon-nucleon cross section. The correlation between the
multiplicity $N_{imf}$ of
intermediate mass fragments and the total numer of charged particles
$N_c$ has the behavior similar to $Q_{zz}$,  which can be used as a
complementary probe to the in-medium nucleon-nucleon cross section.
\end{minipage}
\end{center}
\vskip 0.15in
{\bf PACs Number(s)}: 25$\cdot$70$\cdot$pg, 02$\cdot$70$\cdot$
Ns, 24$\cdot$10$\cdot$Lx
\newpage
\baselineskip 0.3in
\section{Introduction}
\hskip 0.3in
In recent years, the establishment of radioactive
beam facilities at many laboratories over the world and the use of
radioactive beams with large neutron or proton excess have
offered an excellent opportunity to investigate the
isospin-dependence of heavy ion collision (HIC)
dynamics$^{[1-4]}$. This kind of study has made it possible to
obtain the information about the equation of state (EOS) of
asymmetric nuclear matter ranging from symmetric nuclear matter
to pure neutron matter and  the information on isospin-dependence
of in-medium nucleon-nucleon (N-N) cross section, which are
significantly important not only in  understanding  nuclear
properties but also in
exploring  the explosion mechanism of supernova and the cooling
rate of neutron stars. However two essential ingredients in HIC
dynamics, the symmetry potential of the mean field and the
isospin-dependent in-medium N-N cross section, have not been well
determined so far. Recently, Li et al.$^{[1,5,6]}$ made use of the
isospin dependent transport theory to investigate nuclear
symmetry energy and showed that the rate of pre-equilibrium
neutron-proton emitted in intermediate energy HIC
is sensitive to the density dependence of nuclear
symmetry potential, but insensitive to the incompressibility of
symmetric nuclear matter and the in-medium N-N cross section. Also,
R.Pak and Bao-An Li et al. have found that the isospin dependence
of collective flow and balance energy are mainly originated from
the isospin-dependent in-medium N-N cross section$^{[5],[7-11]}$.
Recently we have found that nuclear stopping can be used as a good
probe for exploring the in-medium N-N cross section in HIC in
the beam energy ranging from above Fermi energy to 150MeV/nucleon.
However, it is still not clear how the nuclear stopping
depends on the neutron-proton ratio
of the colliding system$^{[12]}$. In viewing that little
information is known about the in-medium N-N cross
section and its isospin dependence up to now, it is thus very
desirable to find an efficient way to gain such kind of knowledge.
\par
It is known that many intermediate mass fragments(IMF) are emitted in the
process of intermediate energy HIC and that the element distribution
and fragment multiplicity indicate a strong correlation between the
multiplicity of intermediate mass fragments $N_{imf}$ and the total
number of charged particles $N_{c}$ $^{[7]}$. The question is
whether the multifragmentation and dissipation, especially the
$N_{imf}$ and $Q_{zz}$ are sensitive to the neutron-proton ratio
of the colliding system. To answer this question, we have
investigated the isospin effect on the process of multifragmentation
and dissipation in HIC for the colliding systems with
different ratios of neutron to proton  by using isospin dependent
quantum molecular dynamics (IQMD). To increase the efficiency of
detectors and the statistics of $N_{imf}$ and $Q_{zz}$, the
reversible kinetic reactions with heavy projectiles on light
targets are suggested to have more intermediate
mass fragments emitted towards forward angles.
\par
The calculated results show prominent isospin effect for the
multifragmentation $N_{imf}$ and the quadrupole of single particle
momentum distribution $Q_{zz}$ for the colliding systems with
different neutron or proton excesses. The multiplicity of
intermediate mass fragments $N_{imf}$ of neutron-poor systems is
always larger than that of neutron-rich systems,
while the quadrupole of single particle momentum distribution
$Q_{zz}$ for neutron-poor systems is smaller than that of
neutron-rich systems at the beam energies from 50MeV/nucleon
to 150MeV/nucleon for all the reversible colliding systems
studied here.
The mechanism responsible for the above isospin effect can be found
from the fact that the mean N-N cross section for a neutron-poor
system is larger than that of the corresponding neutron-rich system
with the same masses of projectile and target. The calculated results
also show that the correlation between the multiplicity of
intermediate mass fragments $N_{imf}$ and the total number of
charged particles $N_{c}$, depends strongly on the isospin
dependence of in-medium N-N cross section and weakly on the
symmetry potential in the chosen beam energy region.
In Ref.[12], it is found that $Q_{zz}$ is very sensitive to the
isospin dependence of in-medium N-N cross section and insensitive
to the symmetry potential.
In the present paper, the calculations show that the conclusion
drawn in Ref.[12] about $Q_{zz}$ remains true for the reversible
kinetic reactions with heavy projectiles on light targets, but
it is found that $Q_{zz}$ increases slightly as increasing the
neutron-proton ratio of the colliding system. Therefore, $Q_{zz}$
can be a good probe for extracting information on in-medium N-N
cross section in HIC in the beam energies ranging from above Fermi
energy to 150MeV/nucleon$^{[12]}$, while the correlation between
$N_{imf}$ and $N_{c}$ may serve as a complementary one.
\par
\section{Theoretical Model}
\hskip 0.3in
The dynamics of intermediate energy HIC described by QMD$^{[13-14]}$
contains three ingredients: density dependent mean field,
in-medium N-N cross section and Pauli blocking. To describe isospin
effect appropriately, QMD should be modified properly:
the density dependent mean field should contain correct
isospin terms including symmetry energy and Coulomb potential,
the in-medium N-N cross section should be different for
neutron-neutron ( proton-proton ) and neutron-proton collisions,
and the Pauli blocking should be counted by distinguishing neutrons
and protons. In addition, the initial condition of the ground state
of two colliding nuclei should also contain isospin information.
In the present calculation, the ground state of each
colliding nucleus is prepared by using the initial code of IQMD,
according to its density distribution
obtained from the Skyrme-Hatree-Fock calculation with the
parameter set of SKM$^{*}$$^{[15]}$.
The interaction potential
is as follows,
\begin{equation}
U(\rho)=U^{Sky}+V_{c}(1-\tau_{z})+U^{sym}+V^{yuk}+U^{MDI}+U^{Pauli}
\end{equation}
$U^{Sky}$ is the density-dependent Skyrme potential,
\begin{equation}
U^{Sky}=\alpha (\frac \rho {\rho _0})+\beta (\frac \rho {\rho _0})^\gamma
\end{equation}
$V_{c}$ is Coulomb potential. $U^{Yuk}$ is the Yukawa potential$^{[13]}$ ,
\begin{equation}
\begin{array}{ll}
U_{j}^{Yuk}= & t_3\sum_{i \neq j}\frac{e^{L/m^{2}}}{r_{ij}/2m}\{
e^{-r_{ij}/m}[1-\Phi(\sqrt{L}/m-r_{ij}/2\sqrt{L})]-\\
&\\
 &e^{r_{ij}/m}[1-\Phi(\sqrt{L}/m+r_{ij}/2\sqrt{L})]\}
\end{array}
\end{equation}
where $\Phi$ is the error function.
$U^{MDI}$ is the momentum dependent interaction ( MDI )$^{[16]}$,
\begin{equation}
U^{MDI}=t_4ln^2[t_5(\overrightarrow{p_1}-\overrightarrow{p_2})^2+1]\frac
\rho {\rho _0}
\end{equation}
$U^{Pauli}$ is the Pauli potential$^{[17-18]}$,
\begin{equation}
U^{Pauli}=V_p\{\frac \hbar {p_0q_0})^3exp(-\frac{(\overrightarrow{r_i}-
\overrightarrow{r_j})^2}{2q_0^2}-\frac{(\overrightarrow{p_i}-\overrightarrow{
p_j})^2}{2p_0^2}\}\delta _{p_ip_j}
\end{equation}
 $$\delta_{p_{i}p_{j}}=\left\{ \begin{array}{ll}
              1 & \mbox{for neutron-neutron or proton-proton}\\
              0 & \mbox{for neutron-proton}
             \end{array}
            \right. $$
$U^{sym}$ is the symmetry potential. In the present paper, three different kinds
of $U^{sym}$ have been used$^{[1]}$,
\begin{equation}
U_1^{sym}=cF_1(u)\delta \tau _z
\end{equation}
\begin{equation}
U_2^{sym}=cF_2(u)\delta \tau _z+\frac 12cF_2(u)\delta ^2
\end{equation}
\begin{equation}
U_3^{sym}=cF_3(u)\delta \tau _z-\frac 14cF_3(u)\delta ^2
\end{equation}
 with
    \[\tau_{z}=\left\{ \begin{array}{ll}
              1 & \mbox{for neutron}\\
             -1 & \mbox{for proton}
             \end{array}
            \right. \]
Here $c$ is the strength of the symmetry potential, taking the value of
32MeV. $F_1(u)=u$, $F_2(u)=u^2$ and
$F_3(u)=u^{1/2}$, $u\equiv \frac \rho {\rho_0}$, $\delta $
is the relative neutron excess $\delta =\frac{\rho _n-\rho _p}
{\rho _n+\rho _p}=\frac{\rho _n-\rho _p}\rho$.
$\rho$ and $\rho_0$, $\rho_n$ and $\rho_p$ are the nuclear density and
its normal value, neutron density and and proton density, respectively.
The parameters of the interaction potentials are given in table 1.
\begin{center}
 Table 1. The parameters of the interaction potentials
\vskip 0.3in
\begin{tabular}{|c|c|c|c|c|c|c|c|c|c|} \hline
\small
$\alpha$ & $\beta$ &$\gamma$&$t_{3}$&m&$t_{4}$&$t_{5}$&$V_{p}$&$p_{0}
$&$q_{0}$\\ \hline
(MeV)&(MeV)&&(MeV)&(fm)&(MeV)&($MeV^{-2}$)&(MeV)&(MeV/c)&(fm)\\\hline
$-$390.1&320.3&1.14&7.5&0.8&1.57&$5\times10^{-4}$&30&400&5.64\\\hline
\end{tabular}
\end{center}
\vskip 0.3in
It is worth mentioning that recent studies of collective flow in HIC
at intermediate energies have indicated a reduction of the in-medium N-N
cross sections$^{[19-21]}$. An empirical expression of the in-medium
N-N cross section$^{[21]}$ is used:
\begin{equation}
\sigma^{med}_{NN}=(1+\alpha\frac{\rho}{\rho_{0}})\sigma^{free}_{NN}
\end{equation}
with the parameter $\alpha\approx-0.2$ which has been found to better
reproduce the flow data $^{[19-20]}$. Here $\sigma^{free}_{NN}$ is the
experimental N-N cross section$^{[22]}$. The neutron-proton cross
section is about 3 times larger than the proton-proton or
neutron-neutron cross section below 300 MeV.
\par
$Q_{zz}$ will be used to describe the nuclear stopping:
\begin{equation}
Q_{ZZ}=\sum_{i}^{A} (2P_{z}(i)^{2}-P_{x}(i)^{2}-P_{y}(i)^{2}).
\end{equation}
\par
In order to make the isospin effect on the multifragmentation
process in HIC more prominent, comparable study is carried out for
two pairs of reversible reaction systems. For each pair of comparable
reaction systems the same mass of heavy projectiles and light targets,
the same incident energy and the same impact parameter have been used
as follows: $^{120}_{48}Cd+^{40}_{18}Ar$ and
$^{120}_{54}Xe+^{40}_{20}Ca$ with the neutron-proton ratios
1.42 and 1.16;
$^{76}_{30}Zn+^{40}_{18}Ar$ and $^{76}_{36}Kr+^{40}_{20}Ca$
with the neutron-proton ratios 1.42 and 1.07.
Hence the differences of $N_{imf}$ and $Q_{zz}$ for each pair of the
comparable colliding systems are mainly due to the isospin effect on
the process of multifragmentation and dissipation.
\par
By means of the modified isospin-dependent coalescence model$^{[23]}$,
we construct clusters within which the particle relative momentum is
smaller than $p_{0}$= 300MeV/c and the relative distance is smaller
than $R_{0}$= 3.5fm. To avoid the nonphysical clusters,
the restructured aggregation model$^{[24]}$ is used until no
nonphysical cluster is produced.
\section{Results and Discussions}
\hskip 0.3in
 $N_{imf}$ and $N_{c}$ are calculated event by event with the
charge number of IMF from Z=3 to 13 for  the heavy
colliding systems, and Z=3 to 8 for  the medium mass colliding
systems.
\subsection{The isospin effect on the process of multifragmentation and
dissipation in HIC}
\hskip 0.3in
Fig.1 shows the time evolution of $N_{imf}$ for the central
reactions $^{76}Zn+^{40}Ar$ (line 1) and $^{76}Kr+^{40}Ca$ (line 2)
at $E$=80 MeV/nucleon (left panel), and  $^{120}Cd+^{40}Ar$ (line 1)
and $^{120}Xe+^{40}Ca$ (line 2) at $E$=100 MeV/nucleon (right panel).
From Fig.1 one can see that $N_{imf}$ for the two neutron-poor colliding
systems $^{120}Xe+^{40}Ca$ and $^{76}Kr+^{40}Ca$ are larger than
those for the neutron-rich systems
$^{120}Cd+^{40}Ar$ and $^{76}Zn+^{40}Ar$.
This difference are mainly due to the isospin effect on the
multifragmentation because other reaction conditions, except the
ratios of neutron to proton, are the same.
\par
As is well known that nuclear reaction mechanism and reaction yields are
sensitive to both impact parameter and incident energy. Denoted by the same
lines  as in Fig.1,  Fig.2  depicts the
multiplicity of intermediate mass fragments $N_{imf}$ as a function
of impact parameter for the two pairs of colliding systems,
$^{120}_{48}Cd+^{40}_{18}Ar$ and $^{120}_{54}Xe+^{40}_{20}Ca$,
$^{76}_{30}Zn+^{40}_{18}Ar$ and $^{76}_{36}Kr+^{40}_{20}Ca$.
The incident energy $E$ is 100 MeV/nucleon for the heavy
systems (left panel), and 80 Mev/nucleon  for the medium
systems (right panel).
As in Fig.1, at small impact parameters, the lines labeled 2
for the neutron-poor colliding systems are always above the corresponding
lines labeled 1 for the neutron-rich systems.
The difference between them disappears gradually as increasing
impact parameter.
\par
Fig.3 shows $N_{imf}$ as a function of beam energy from
15 MeV/nucleon to 200 MeV/nucleon at $b$=0 fm for the above two pairs of
colliding systems with the line labels as  in Fig.1.
The relative locations between the two lines in each panel
of the figure are also similar to those in Fig.1, namely,
$N_{imf}$ for the two neutron-poor colliding systems
$^{120}Xe+^{40}Ca$ and $^{76}Kr+^{40}Ca$ are larger than
those for the corresponding neutron-rich systems $^{120}Cd+^{40}Ar$
and $^{76}Zn+^{40}Ar$ for the beam energy above 50MeV/nucleon.
However, as the beam energy decreases to below  50MeV/nucleon
the collision dynamics is governed by both the mean field and
the nucleon-nucleon collisions.
In this case, the difference between the two lines of the
colliding systems in each pair vanishes gradually.
\par
From Figs.1, 2, and 3 we can see that the intermediate mass
fragment multiplicity $N_{imf}$ for the neutron-poor colliding
system are larger than that of the corresponding neutron-rich system
with the same mass projectile on the same mass target and with the same
entrance channel conditions except that the ratios of neutron to
proton of the colliding systems are different.
\par
The correlation between $N_{imf}$ and $N_{c}$ for the central
collision of the above two pairs of colliding systems with the same
beam energies as above is shown in Fig.4. It is seen from the figure
that the correlation between $N_{imf}$ and $N_{c}$ for the two pairs of
colliding systems displays a clear isospin effect, i.e.,
the $N_{imf}$-$N_{c}$ correlation for the neutron-poor systems is
different from that of the neutron-rich systems. Considering
the total yield of  $N_{imf}$ (namely the integral area of each
curve in the figure), one can reach the same conclusion as from Fig.1,
Fig.2, and Fig.3. This means that the $N_{imf}$ for neutron-poor
systems (solid line), on average, is larger  than that for the
corresponding neutron-rich systems (dash line). The difference
comes mainly  from the isospin effect on multifragmentation  in
intermediate energy HIC.
\par
The mechanism of the above fragmentation process can be
understood as follows. Experimentally the free neutron-proton cross
section is about three times larger than the free neutron-neutron or
proton-proton cross section below 300 MeV. The mean N-N cross section
is defined as:
\begin{equation}
<\sigma>=\frac{N_{np}\sigma_{np}+(N_{nn}+N_{pp})\sigma_{pp}}{N_{np}+N_{pp}
+N_{nn}}=\frac{(3N_{np}+N_{nn}+N_{pp})}{N_{np}+N_{nn}+N_{pp}}\sigma_{pp}
=(1+\frac{2N_{np}}{N})\sigma
\end{equation}
where $N_{np}$, $N_{nn}$ and $N_{pp}$ are the collision numbers
for neutron-proton, neutron-neutron, and proton-proton,
respectively, and $N= N_{np}+N_{nn}+N_{pp}$. $\sigma_{np}$,
$\sigma_{nn}$, and $\sigma_{pp}$ are the free N-N cross sections
for neutron-proton, neutron-neutron, and proton-proton,
respectively. In general, $\sigma_{nn}=\sigma_{pp}=\sigma$.
Because the total collision numbers for two colliding systems
with the same mass projectile and the same mass target are the same,
and the number of neutron-proton collisions for
neutron-poor colliding system is larger than that of the
neutron-rich colliding system, the mean total N-N cross section
$<\sigma>$ of neutron-poor system is thus larger than that of
neutron-rich system.
Due to the effect of Pauli blocking, the effective
collision numbers become smaller. But after considering
the Pauli blocking, the above conclusion remains unchanged.
Therefore, the neutron-poor system will have more effective
N-N collisions and lead to stronger compression-expansion,
resulting  in a large number of multifragmentation $N_{imf}$
for the neutron-poor system in comparison to  the neutron-rich system
in the above 50MeV region where the N-N collisions are dominant.
\par
In Fig.5 is plotted the time evolution of the quadrupole of
single particle momentum distribution $Q_{zz}$ for the reactions
$^{76}Zn+^{40}Ar$ ( dash line ) and $^{76}Kr+^{40}Ca$ ( solid line )
( bottom  panel), $^{120}Cd+$ $^{40}Ar$ ( dash line )
and $^{120}Xe+$ $^{40}Ca$ ( solid line)(top panel) at
$E$=80, 100, 150MeV/nucleon and $b=0.0$fm.
In the figure, $Q_{zz}$ for the neutron-poor system is always
smaller than that of the corresponding neutron-rich system.
Smaller $Q_{zz}$ indicates larger dissipation  of the
initial longitudinal collective motion into the internal chaotic
motion  and the subsequent thermalization of the system.

\subsection{A good probe and a complementary probe to in-medium
N-N cross section}
\hskip 0.3in
Since the reaction dynamics of HIC is mainly
governed by both nuclear EOS and in-medium N-N cross section,
to understand the collision dynamics in details, both ingredients
should be studied carefully. As is well known that the effects of both
ingredients are usually mixed in the dynamics  and
the main uncertainty of the information about the nuclear EOS
extracted from HIC is due to our poor knowledge of the N-N cross
section in medium. If one can find an experimental probe which can
distinguish the contribution of EOS from that of in-medium N-N cross
section, that will be very desirable.
In this paper we have found that $Q_{zz}$ may serve as a good
probe to in-medium N-N cross section.

In Fig.6 is given $Q_{zz}$ as a function of neutron-proton ratio for
seven colliding systems $^{76}Kr+^{40}Ca$, $^{120}Xe+^{40}Ca$,
$^{64}Ni+^{40}Ar$, $^{86}Kr+^{40}Ar$, $^{76}Zn+^{40}Ar$,
$^{85}Ge+^{40}Ar$, $^{74}Ni+^{47}Ar$ with the neutron-proton
ratios 1.07, 1.16, 1.26, 1.33, 1.42, 1.5 and 1.56 at
$E$=100MeV/nucleon and $b$=0.0 fm for four cases:
(1) the symmetry potential $U_1^{sym}$ being employed
and in-medium N-N cross section $\sigma _{NN}^{med}$ being
isospin-dependent, namely, $U_1^{sym}$+$\sigma ^{iso}$;
(2) $U_1^{sym}$ and N-N cross section $\sigma _{NN}^{med}$ being
isospin-independent, denoted by $U_1^{sym}$ + $\sigma ^{noiso}$;
(3) $U_2^{sym}$ and $\sigma _{NN}^{med}$ being isospin-dependent,
denoted by U$_2^{sym}$+ $\sigma ^{iso}$;
(4) $U_3^{sym}$ and $\sigma _{NN}^{med}$ being isospin-dependent,
denoted by $U_3^{sym}$+ $\sigma ^{iso}$.
In fig.6, lines labeled 1, 2, 3 and 4 correspond to the above four
cases.

It is clear to see that $Q_{zz}$ depends strongly on the isospin
dependence of in-medium N-N cross section and weakly on the symmetry
potential ( namely, line 1 is located near lines 3 and 4, but far
away from line 2 in Fig.6), though $Q_{zz}$ increases slightly with
increasing neutron-proton ratio of asymmetry colliding system. In
this case, $Q_{zz}$ is a good probe for extracting information on the
isospin dependence of in-medium N-N cross section. We have discussed
this in more details in Ref.[12], but it is not clear in Ref.[12] how
$Q_{zz}$ depends on the neutron-proton ratio of colliding system.
In addition, we shall report a complementary probe for extracting the
information on in-medium N-N cross section: the correlation
between the multiplicity of intermediate mass fragments $N_{imf}$
and the total number of charged particles $N_{c}$, based on the fact
that it is also sensitive to the in-medium N-N cross section and
insensitive to the symmetry potential in the chosen energy region.
To study the contributions to $N_{imf}$ from different ingredients
separately, we consider four cases as the same as in Fig.6.
\par
In Fig.7 is plotted the time evolution of $N_{imf}$ for two different
asymmetric colliding systems,
$^{76}Zn+^{40}Ar$ at $E=80$MeV/nucleon and $b=0.0$fm (left panel),
$^{120}Xe+$ $^{40}Ca$  at $E=100$MeV/nucleon and $b=0.0$fm (right panel).
It is noted that lines 1 are always located near lines 3 and 4, but far
away from lines 2 with increasing colliding time. This implies that
the isospin effect of in-medium N-N cross section on $N_{imf}$
is more important than that of the symmetry potential in the energy
region studied here.
\par
As is well known that nuclear reaction products and reaction
mechanism sensitively depend on impact parameter. Fig.8 shows the
multiplicity of intermediate mass fragments $N_{imf}$ as a
function of impact parameter for the four cases as illustrated in
Fig.6  for the reactions $^{76}Zn+^{40}Ar$
at $E=80$MeV/nucleon ( left panel ) and $^{120}Xe+^{40}Ca$ at
$E=100$MeV/nucleon ( right panel ).
As in Fig.6, lines 1 in the figures locate near lines 3 and 4, but
far away from lines 2 at small impact parameters. This indicates
again that the isospin effect of in-medium N-N cross section on
$N_{imf}$ is larger in comparison with that of symmetry
potential. With increasing impact parmeter this isospin effect
disappears gradually.
\par
The correlations between $N_{imf}$ and $N_{c}$ for the four cases
are plotted in Fig.9 for the reactions
$^{76}Zn+^{40}Ar$ at $E=80$MeV/nucleon ( left panel )
and $^{120}Xe+^{40}Ca$ at $E=100$MeV/nucleon ( right panel ).
The behaviors of the four kinds of lines in each panel are similar
to those in Fig.6, namely, the integral area of the curve 1 is
always close to those of the curves 3 and 4, but larger than that
of the curve 2. From Fig.7, Fig.8, and Fig.9 one can see that
the isospin effect of in-medium N-N cross section on the correlation
between $N_{imf}$ and $N_{c}$ is more important than that of symmetry
potential. Here it should be stressed that we have to choose the
incident energy for certain asymmetric colliding system carefully
to get the above feature, namely, in this energy region the
N-N collisions should be dominant.
In this case we may conclude that the correlation between the
multiplicity of intermediate mass fragments $N_{imf}$ and the
total number of charged particles $N_{c}$ can be used as a
complementary probe to the isospin dependence of the in-medium
N-N cross section in HIC.
\section{Summary and conclusions}
\hskip 0.3in
Starting from the simulation of the intermediate energy HIC by
using IQMD, the calculated results have shown prominent
isospin effects on the process of multifragmentation and dissipation,
i.e., the intermediate mass fragment multiplicity $N_{imf}$
for a neutron-poor colliding system is always larger than that of
the corresponding neutron-rich system, while the quadrupole of
single particle momentum distribution $Q_{zz}$ for a neutron-poor
system is smaller than that of the neutron-rich system with the
same masses of projectile and target, and the same entrance channel
conditions for small impact parameters.
We also can see that $Q_{zz}$ increases slight with increasing
ratio of neutron to proton in the colliding system.
\par
From the theoretical simulation, it is clear to see that $Q_{zz}$
depends strongly on the isospin dependence of in-medium N-N cross
section and weakly on the symmetry potential in the beam energies
ranging from about 50 MeV/nucleon to 150MeV/nucleon.
And the correlation between $N_{imf}$ and $N_{c}$ has the same
properties as $Q_{zz}$ in the chosen energy region. So we would
suggest that $Q_{zz}$ may serve as a good probe and $N_{imf}$ a
complementary probe for extracting the information on the isospin
dependent N-N cross section in HIC.
\section*{Acknowledgment}
\hskip 0.3in This work was supported in part by 973 Project Grant
No.G2000077400, 100 person Project of the Chinese Academy of Sciences,
the National Natural Foundation of China under Grants No. 19775057
and No. 19775020, No. 19847002, No. 19775052 and
KJ951-A1-410, and by the Foundation of the Chinese Academy of Sciences

\newpage
\baselineskip 0.2in
\begin{figure}
\centerline{\epsfig{figure=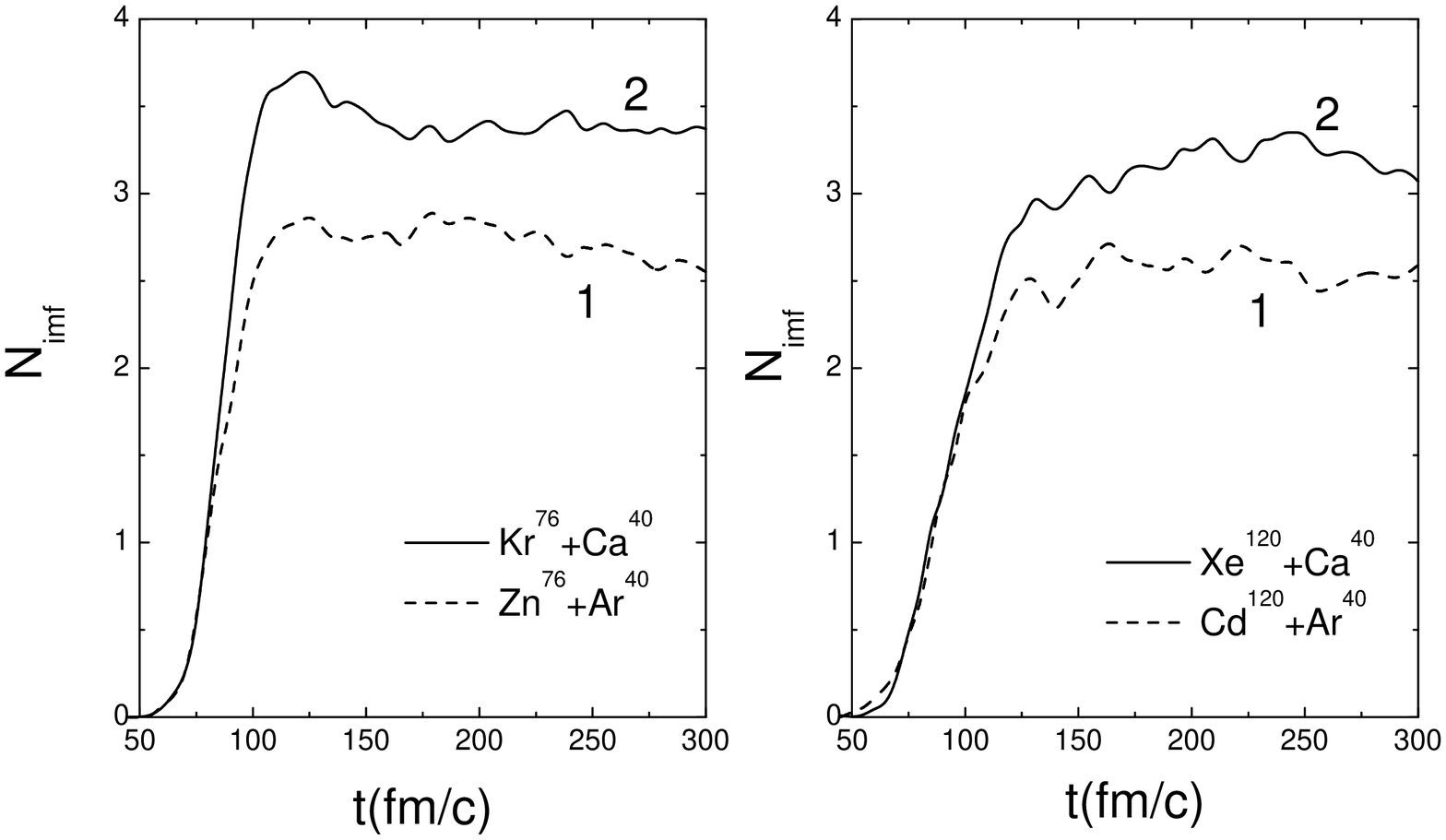,width=14cm}} \caption{The
time evolution of $N_{imf}$ for the central collisions
$^{120}Cd+^{40}Ar$ (line 1) and $^{120}Xe+^{40}Ca$ (line 2) at
$E=100$MeV/nucleon (right panel), $^{76}Zn+^{40}Ar$ (line 1) and
$^{76}Kr+^{40}Ca$ (line 2) at $E=80$MeV/nucleon (left panel).}

\centerline{\epsfig{figure=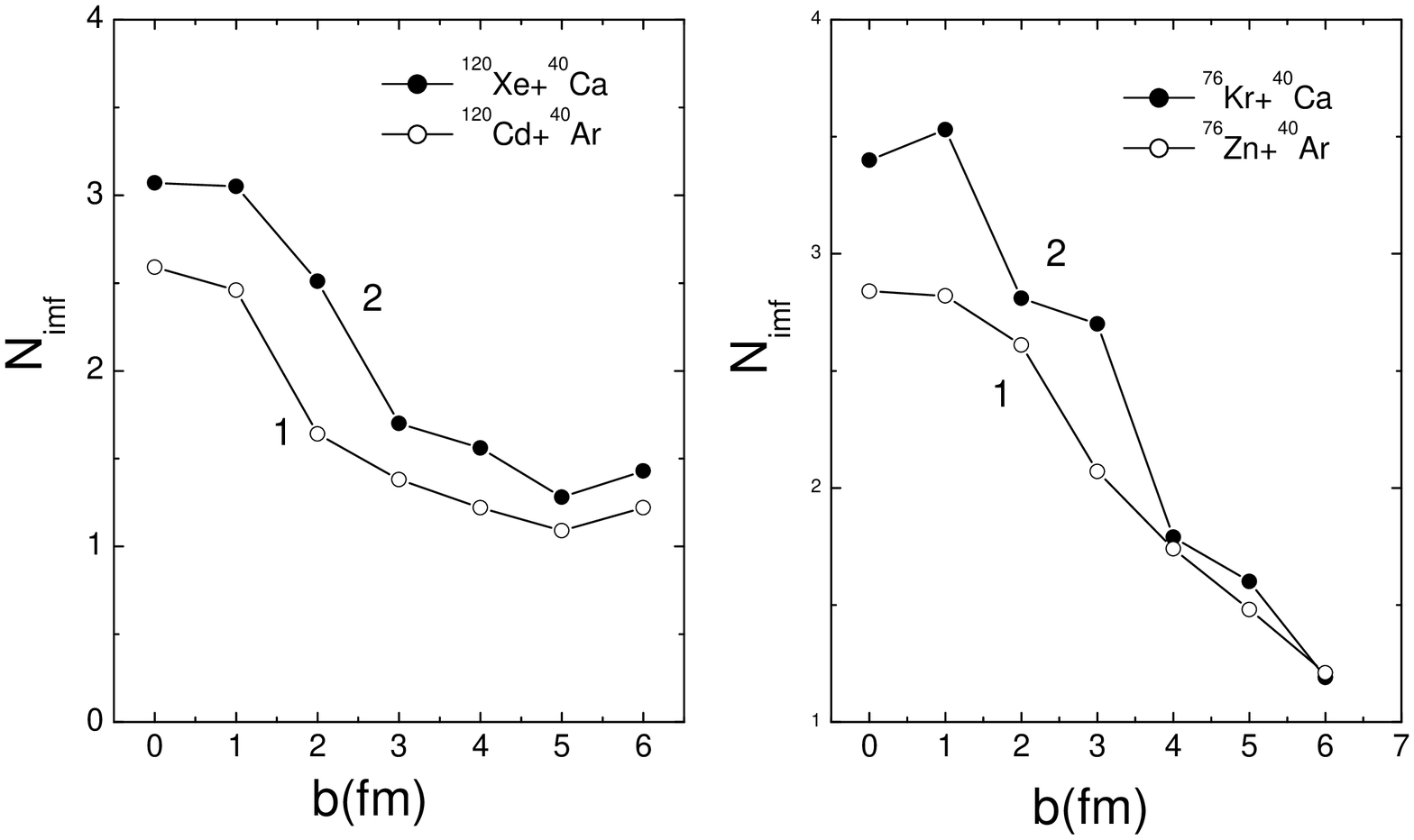,width=14cm}} \caption{ The
multiplicity of intermediate mass fragments $N_{imf}$ as a
function of impact parameter for the reactions $^{120}Cd+^{40}Ar$
( line 1 ) and $^{120}Xe+^{40}Ca$ ( line 2 ) at $E=100$MeV/nucleon
( left panel ), and the reactions $^{76}Zn+^{40}Ar$ ( line 1 ) and
$^{76}Kr+^{40}Ca$ ( line 2 ) at $E=80$MeV/nucleon ( right panel).}
\end{figure}

\newpage
\begin{figure}

\centerline{\epsfig{figure=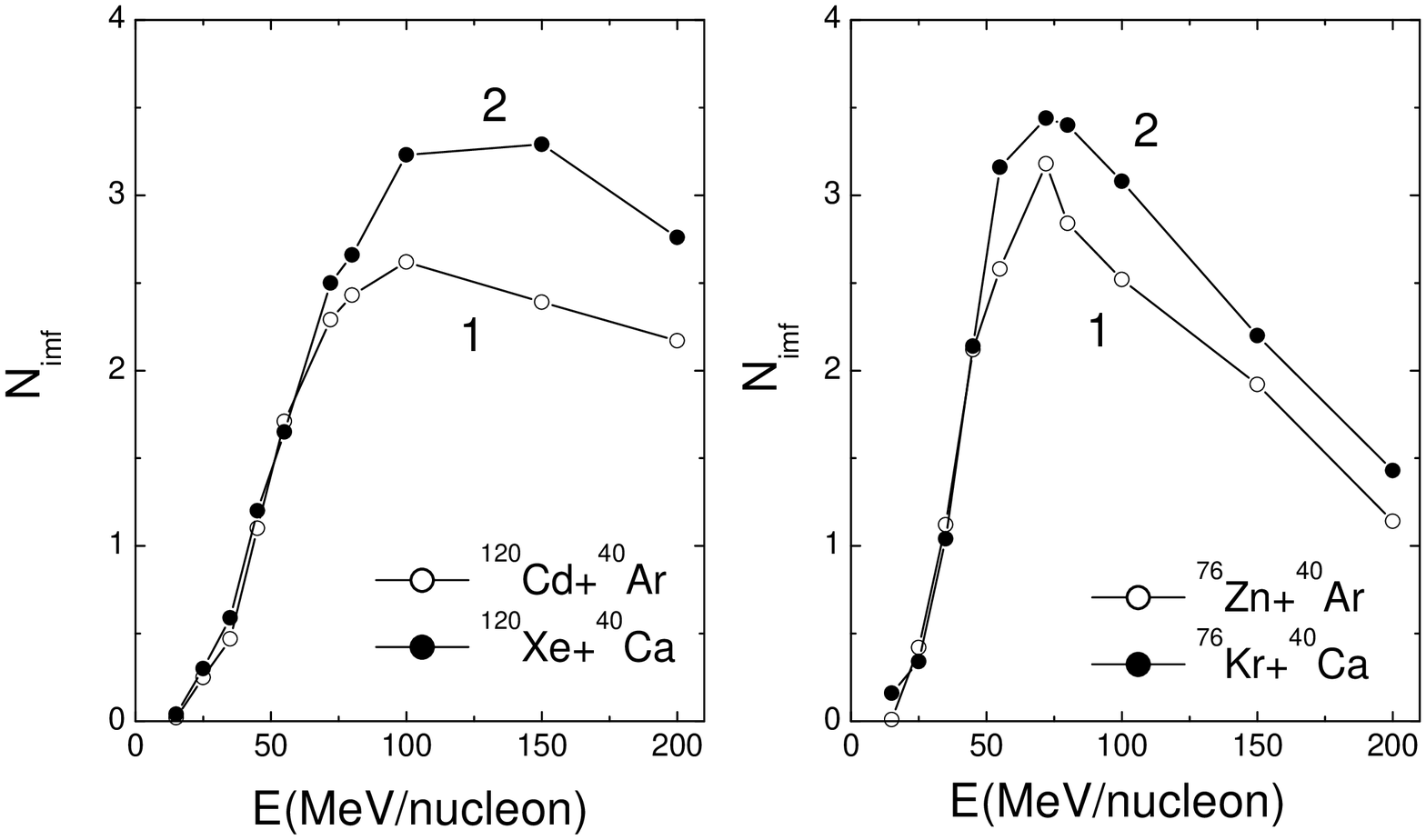,width=14cm}} \caption{ The
$N_{imf}$ as a function of incident energies from 15MeV/nucleon to
200MeV/nucleon at $b$=0 fm for the reactions $^{120}Cd+^{40}Ar$ (
line 1 ) and $^{120}Xe+^{40}Ca$ ( line 2 ) at $E=100$MeV/nucleon
( left panel ), and the reactions $^{76}Zn+^{40}Ar$ ( line 1 )
and $^{76}Kr+^{40}Ca$ ( line 2 ) at $E=80$MeV/nucleon ( right
panel ). }

\centerline{\epsfig{figure=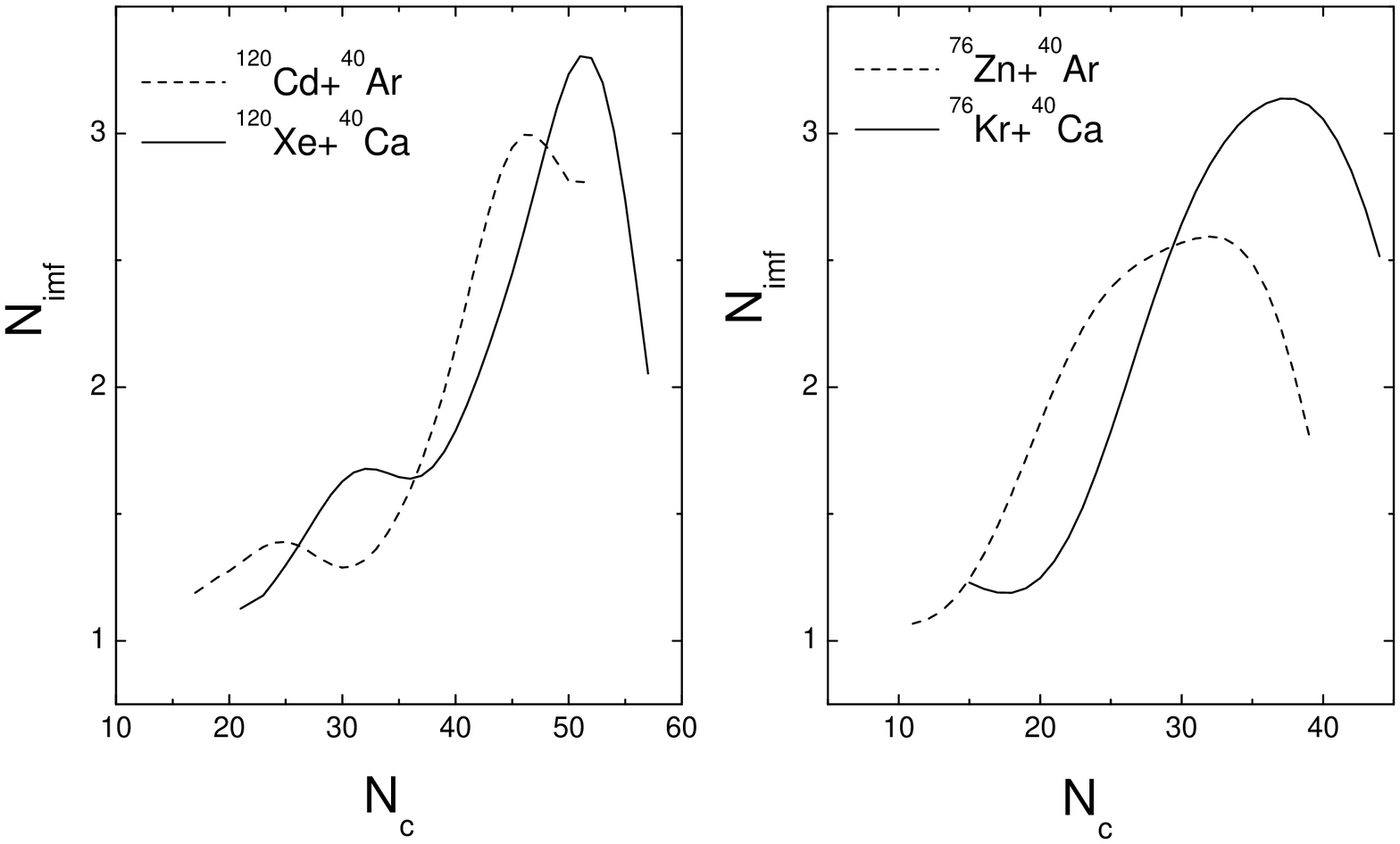,width=14cm}} \caption{ The
correlations between $N_{imf}$ and $N_{c}$ for the reactions
$^{120}Cd+^{40}Ar$ ( dash line ) and $^{120}Xe+^{40}Ca$ ( solid
line ) at $E=100$MeV/nucleon ( left panel ), and the reactions
$^{76}Zn+^{40}Ar$ ( dash line ) and $^{76}Kr+^{40}Ca$ ( solid
line ) at $E=80$MeV/nucleon ( right panel ). }
\end{figure}
\newpage
\begin{figure}
\centerline{\epsfig{figure=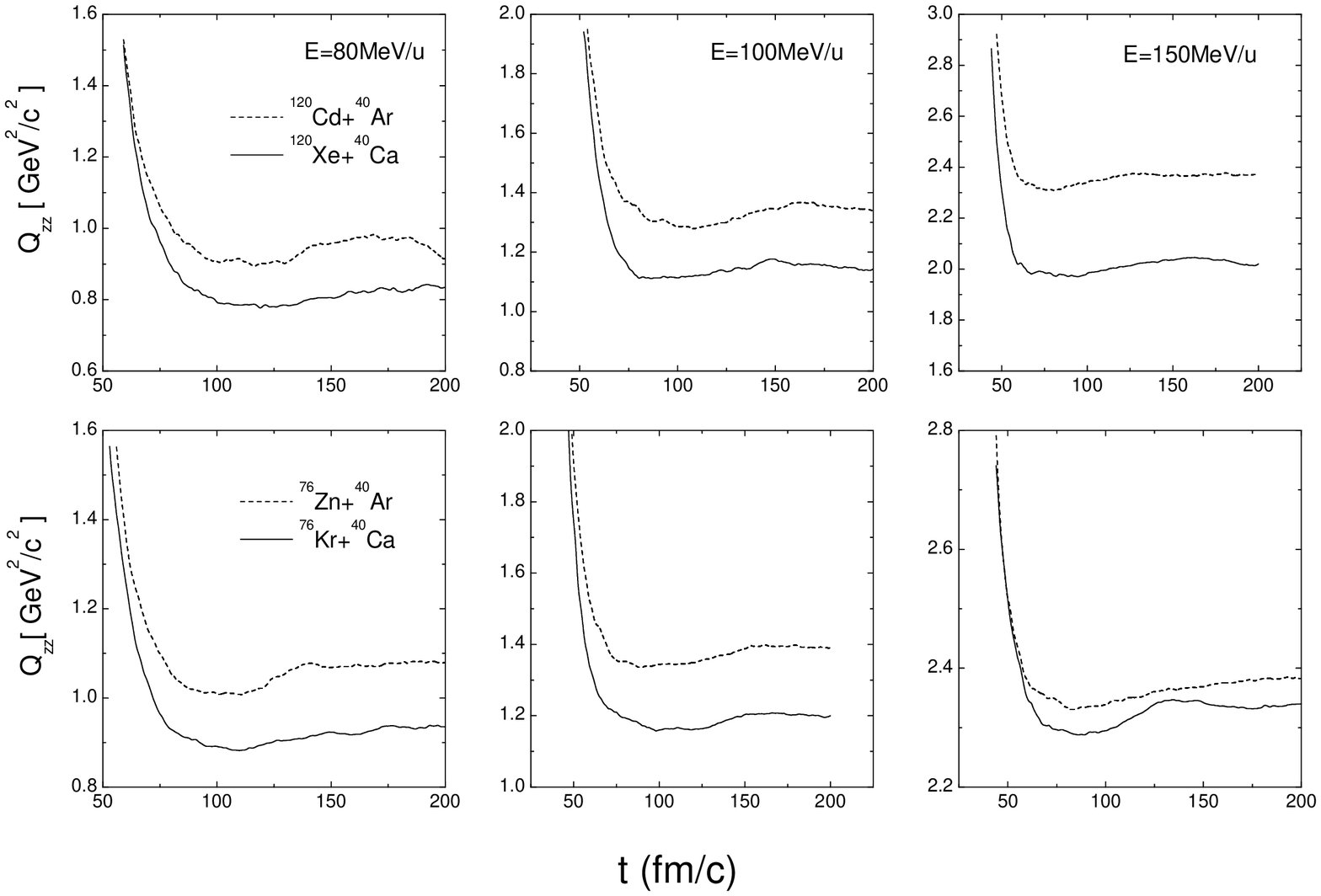,width=12cm}} \caption{ The
time evolution of the quadrupole of single particle momentum
distribution $Q_{zz}$ for the reactions $^{76}Zn+^{40}Ar$ ( dash
line ) and $^{76}Kr+^{40}Ca$ ( solid line ) ( bottom panel ), and
the reactions $^{120}Cd+^{40}Ar$ ( dash line ) and
$^{120}Xe+^{40}Ca$ ( solid line )( top panel ) at $E$=80,100 and
150MeV/nucleon and $b=0.0$fm. }

\centerline{\epsfig{figure=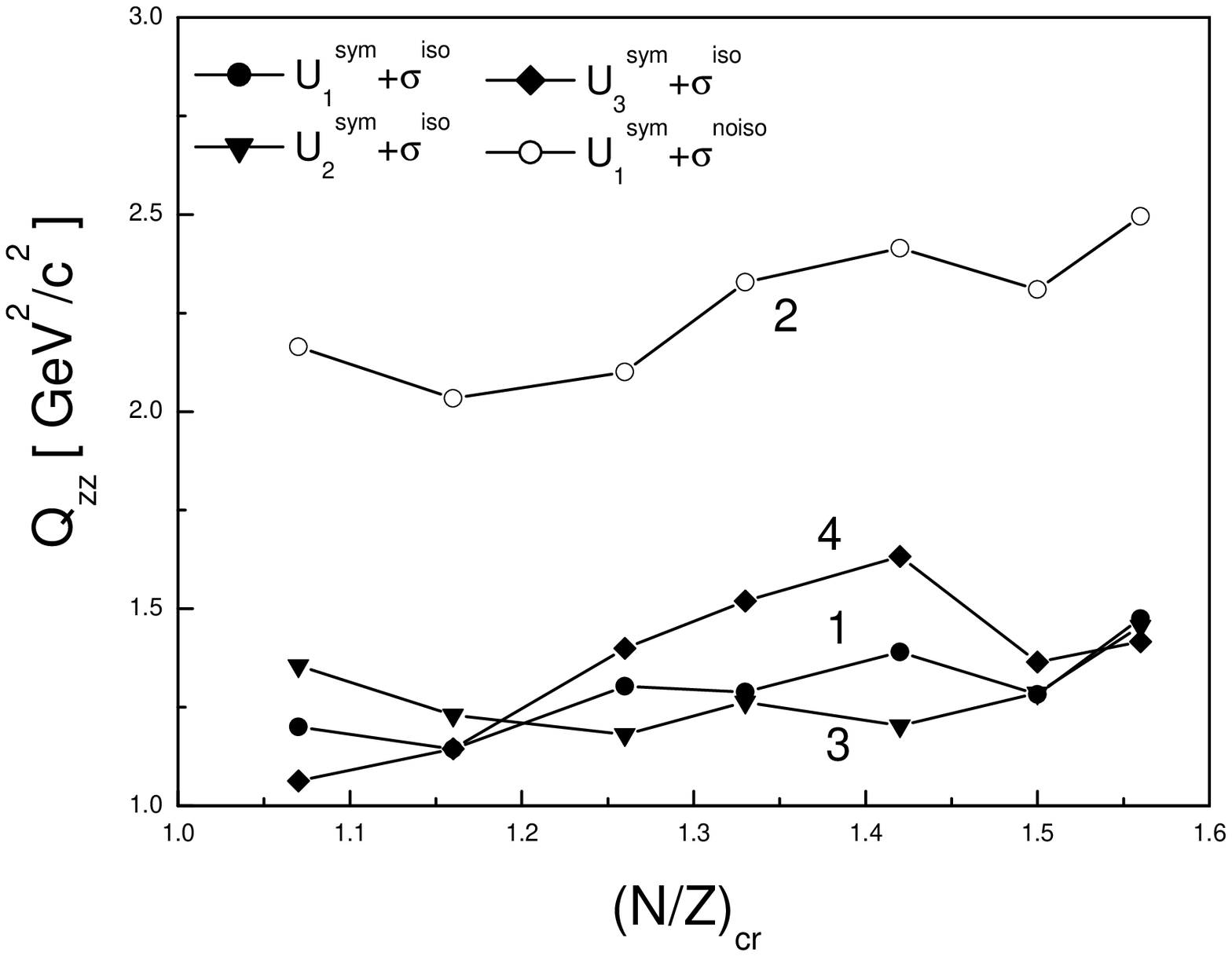,width=12cm}} \caption{ The
quadrupole of single particle momentum distribution $Q_{zz}$ as a
function of the neutron-proton ratio for seven colliding systems
$^{76}Kr+^{40}Ca$, $^{120}Xe+^{40}Ca$, $^{64}Ni+^{40}Ar$,
$^{86}Kr+^{40}Ar$, $^{76}Zn+^{40}Ar$, $^{85}Ge+^{40}Ar$, and
$^{74}Ni+^{47}Ar$ at $E=100$MeV/nucleon and $b=0.0$fm for the four
cases ( see text ). }
\end{figure}
\newpage
\begin{figure}
\centerline{\epsfig{figure=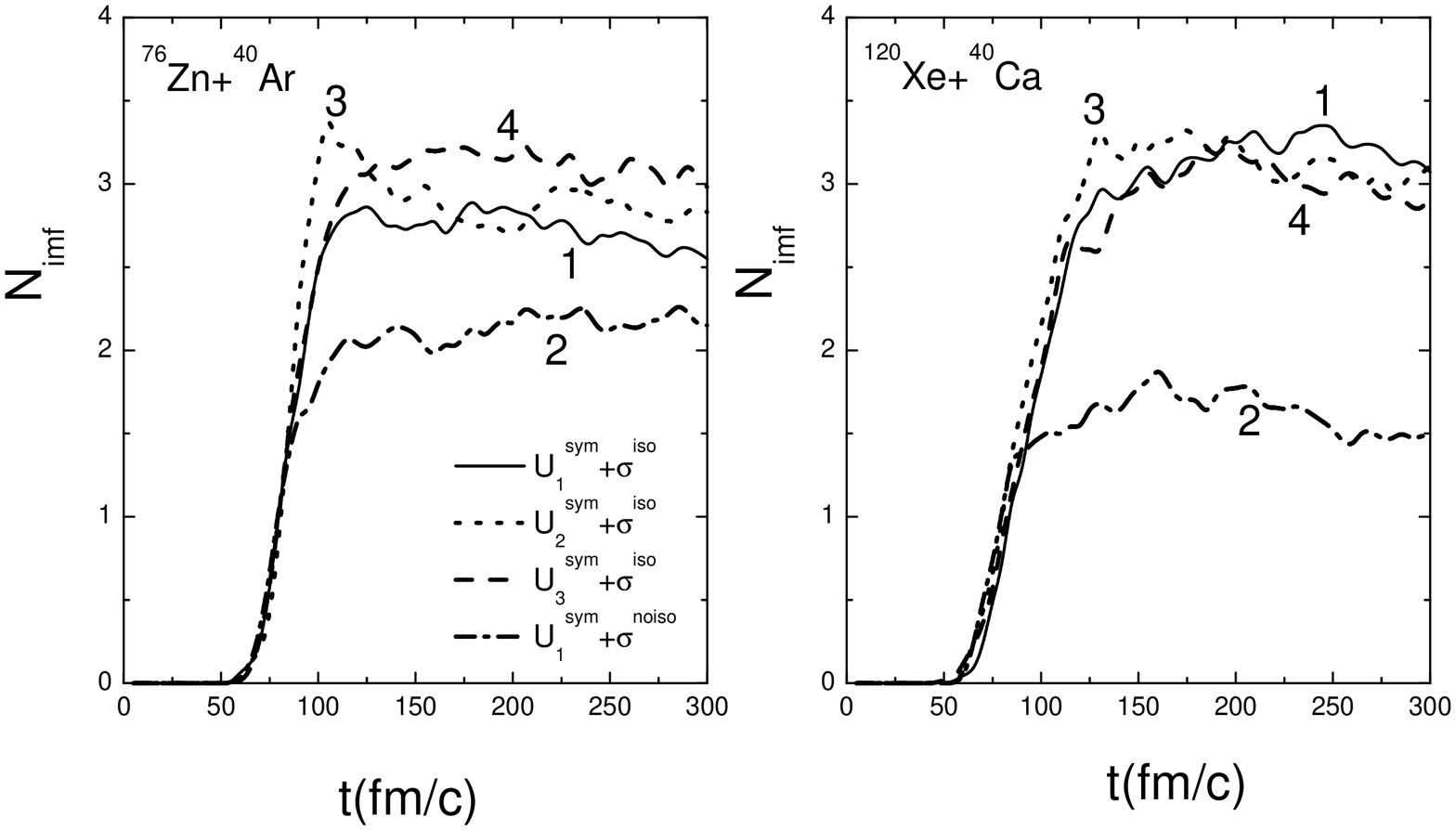,width=14cm}} \caption{ The
time evolution of $N_{imf}$ for the reactions $^{76}Zr+^{40}Ar$ at
$E=80$MeV/nucleon and $b=0.0$fm ( left panel ),
$^{120}Xe+^{40}Ca$ at $E=100$MeV/nucleon and $b=0.0$fm ( right
panel ). Lines 1, 2, 3, and 4 correspond to the four cases  as in
Fig.6. }

\centerline{\epsfig{figure=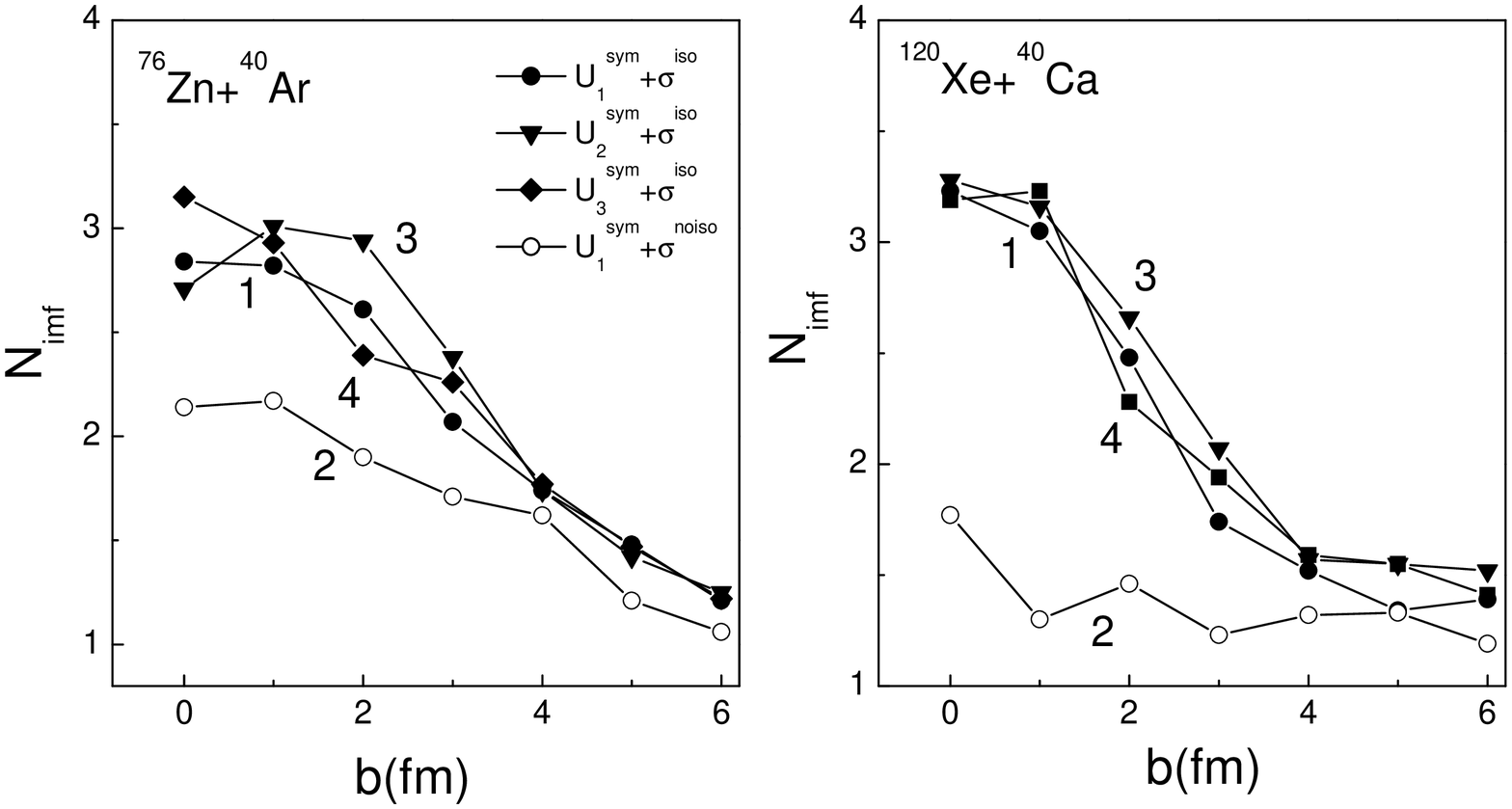,width=14cm}} \caption{
The multiplicity of intermediate mass fragments
$N_{imf}$ as a function of impact parameter for the reactions
$^{76}Zn+^{40}Ar$ at $E=80$MeV/nucleon ( left panel ) and
$^{120}Xe+^{40} Ca$ at $E=100$MeV/nucleon ( right panel ) in the
four cases as
 in Fig.6. }
\end{figure}
\newpage
\begin{figure}

\centerline{\epsfig{figure=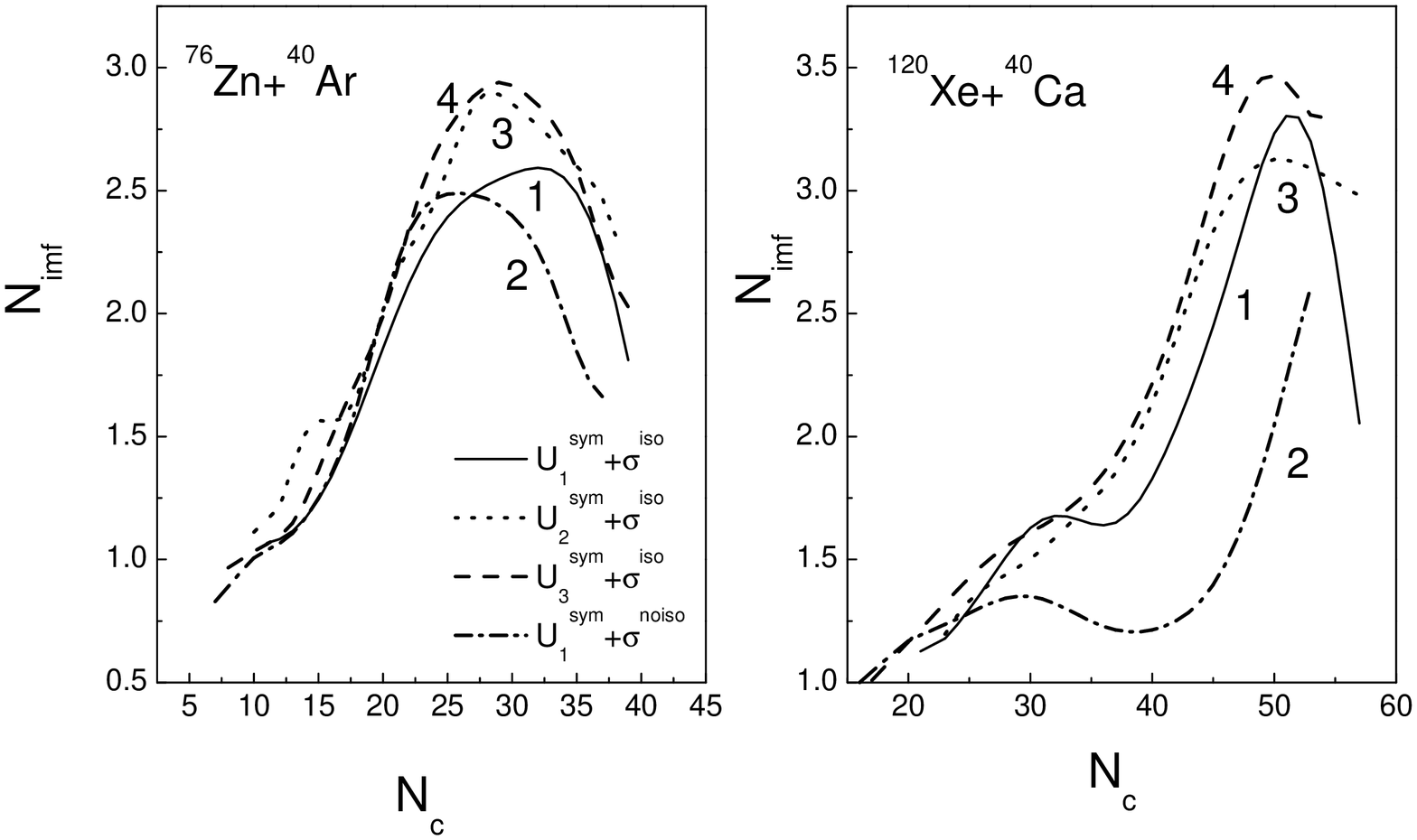,width=14cm}} \caption{ The
correlations between $N_{imf}$ and $N_{c}$ for the reactions
$^{76}Zn+^{40}Ar$ at $E=80$MeV/nucleon ( left panel ) and
$^{120}Xe+^{40}Ca$ at $E=100$MeV/nucleon (right panel) for the
four cases as in Fig.6. }
\end{figure}


\begin{thebibliography}{s29}
\baselineskip 0.28in
\bibitem{s1}Bao-An Li, Che-Ming Ko et al.,
{\it Inter. Jour. Mod. Phys.} {\bf E}(1998) 147-229.
\bibitem{s2}M.Colonna, M.DiToro et al.,
{\it Phys. Rev.} {\bf C57}(1998) 1410-1415.
\bibitem{s3}B. A. Li. Phys. Rev. Lett. 85 (2000)4221-4224.
\bibitem{s4}Bao-An Li and C. M. Ko,
Nucl. Phys. {\it Nucl. Phys.} {\bf A618}(1997)498.
\bibitem{s5}Bao-An Li, Che-Ming Ko and Zhong-zhou Ren et al.,
{\it Phys. Rev. Lett.} {\bf 78}(1997)1644.
\bibitem{s6}Jian-Ye Liu, Qiang Zhao, Shun-Jin Wang, Wei Zuo, Wen-Jun Guo,
{\it Nucl. Phys. A}, to be published.
\bibitem{s7}D. R. Bowman, C. M. Mader et al.,
{\it Phys. Rev.} {\bf C46}(1992)1834.
\bibitem{s8}M. L. Miller, O. Bjarki et al.,
{\it Phys. Rev. Lett.} {\bf 82}(1999)1399-1401.
\bibitem{s9}Yu-Ming Zheng, C. M. Ko, Bao-An Li, Bin Zhang
{\it Phys. Rev. Lett.} {\bf 83}(1999)2534.
\bibitem{s10}R. Pak, W. Benenson, O. Bjarki, J. A. Brown, S. A. Hannuschke,
R. A. Lacey, Bao-An Li, A. Nadasen, E. Norbeck, P. Pogodin, D. E. Russ,
M. Steiner, N. T. B. Stone, A. M. Vander Molen, G. D. Westfull, L. B. Yang,
and S. Y. Yennello, {\it Phys. Rev. Lett.} {\bf 78}(1997) 1022.
\bibitem{s11}R. Pak, Bao-An Li, W. Benenson, O. Bjarki, J. A. Brown,
S. A. Hannuschke, R. A. Lacey, D. J. Magestro, A. Nadasen, E. Norbech,
D. E. Russ, M. Steiner, N. T. B. Stone, A. M. Vander Molen, G. D. Westfall,
L. B. Yang, and S. J. Yennello, {\it Phys. Rev. Lett.} {\bf 78}(1997) 1026.
\bibitem{s12}Jian-Ye Liu, Wen-Jun Guo, Shun-Jin Wang, Wei Zuo, Qiang Zhao.
and Yan-Fang Yang, {\it Phys. Rev. Lett.} {\bf 86}(2001)975.
\bibitem{s13}J. Aichelin, G. Peilert, A. Bohnet, A. Rosenhauer, H. Stocher,
 and W. Greiner, {\it Phys. Rev.} {\bf C37}(1988) 2451.
\bibitem{s14}G. Peilert, H. Stocher, and W. Greiner,
{\it Phys. Rev.} {\bf C39}(1989)1402.
\bibitem{s15}P. G. Reinhard, in Computational Nuclear Physics 1,
eds. K. Langanke, J. A. Maruhn, and S. E. Koonin,
Germany, Springer-Verlag, 1991, P. 28-50.
\bibitem{s16}J. Aichelin, A. Rosenhauer, G. Peilert, H. St$\ddot o$cker
and W. Greiner, {\it Phys. Rev. Lett.} {\bf 58}(1987) 1926.
\bibitem{s17}Hang Liu and Jian-Ye Liu,
    {\it Z. Phys.} {\bf A345}(1996) 311.
\bibitem{s18}C. Dorso, S. Duarte, J. Randrup,
{\it Phys. Lett.} {\bf B188}(1987)287.
\bibitem{s19}M. J. Huang et al., {\it Phys. Rev. Lett.} {\bf 77}(1996)3739.
\bibitem{s20}G. D. Westfall et al.,{\it Phys. Rev. Lett.} {\bf 71}(1993)1986.
\bibitem{s21}D. Klakow, G. Welke and W. Bauer,
{\it Phys. Rev.} {\bf C48}(1993)1982.
\bibitem{s22}K. Chen, Z. Fraenkel et al.,
{\it Phys. Rev.} {\bf 166}(1968)949.
\bibitem{s23}G. F. Bertsch and S. D. Gupta,
{\it Phys. Rep.} {\bf 160}(1988)1991-233.
\bibitem{s24}C. Ngo, H. Ngo and S. Leray et al.,
{\it Phys. Rep.} 499(1989)148.
\end{thebibliography}
\end{document}